\documentclass[aps,prd,preprint,tightenlines,superscriptaddress,showpacs,byrevtex]{revtex4}
 
\usepackage{graphicx} 

\renewcommand{\thefootnote}{\roman{footnote}}

\def\myspecial#1{}                   
\newcommand{\etap}{\eta^{\prime}}
\newcommand{\fz}{f_0}
\newcommand{\az}{a_0}
\newcommand{\ks}{K_S^0}
\newcommand{\kl}{K_L^0}
\newcommand{\dcp}{D_{CP}}
\newcommand{\etac}{\eta_c}
\newcommand{\chic}{\chi_{c0}}

\newcommand{\ul}[1]{\underline{#1}}
\newcommand{\dul}[1]{\underline{\underline{#1}}}
\newcommand{\sinbb}{\sin2\phi_1}

\begin{document}

\preprint{
  \vbox{ 
    \hbox{   }
    \hbox{hep-ph/0402097}
    \hbox{\today}
  }
}

\title{\quad\\[0.5cm] \boldmath
  Time-Dependent $CP$ Violation in $B^0 \to P^0P^0X^0$ Decays
}

\author{Tim Gershon\footnote{\tt gershon@bmail.kek.jp}}
\author{Masashi Hazumi\footnote{\tt masashi.hazumi@kek.jp} }

\affiliation{High Energy Accelerator Research Organization (KEK), Tsukuba}
 
\myspecial{!userdict begin /bop-hook{gsave 280 50 translate 0 rotate
    /Times-Roman findfont 18 scalefont setfont
    0 0 moveto 0.70 setgray
    (\mySpecialText)
    show grestore}def end}

\begin{abstract}
  \noindent
  We note that in decays of the type $B^0 \to P^0P^0X^0$,
  where $P^0$ and $X^0$ represent any $CP$ eigenstate spin-0 neutral particles,
  the final state is a $CP$ eigenstate.
  We consider the possibilities for studying time-dependent $CP$ violation
  in decays of this type at \mbox{$B$ factories} with high luminosity,
  with particular attention to $B^0 \to \ks \ks \ks$.
  We also comment on some cases where $X^0$ has non-zero spin,
  and decays of the type $B^0 \to P^0 Q^0 X^0$,
  where the three final state particles are different spin-0 $CP$ eigenstates.
\end{abstract}


\maketitle

{\renewcommand{\thefootnote}{\fnsymbol{footnote}}}
\setcounter{footnote}{0}

\section{Introduction}

Recent years have seen a large amount of new experimental information
on the phenomena of $CP$ violation, one of the least well understood 
aspects of the Standard Model of particle physics.
The observation of mixing-induced $CP$ violation in 
\mbox{$B^0 \to J/\psi \ks$} decays~\cite{ref:belle_sin2phi1,ref:babar_sin2phi1}
and the observation of direct $CP$ violation in the 
neutral kaon system~\cite{ref:eoe} are both in agreement with the 
predictions of the Kobayashi-Maskawa theory~\cite{ref:km}.
The KM mechanism is thus established as an effective and elegant 
description of the weak interactions of the Standard Model quarks.

Nonetheless, it would be premature to claim that $CP$ violation is understood.
Internationally, a large number of experiments,
at various stages from planning to running, are intended to 
elucidate further information on this most mysterious of phenomena.
Testing the validity of the KM mechanism as the sole source of $CP$ violation
is one of the major experimental challenges in high energy physics today.

Already, there are some tantalizing hints that $CP$ violating observables
may provide evidence for physics beyond the Standard Model.
In particular, the world average for the mixing-induced $CP$ violation
parameter in $b \to s q\bar{q}$ decays 
($q = u,d,s$; {\it e.g.} $B^0 \to \phi \ks$),
is only marginally consistent with that measured in 
$b \to c \bar{c} s$ decays 
({\it e.g.} $B^0 \to J/\psi \ks$)~\cite{ref:belle_phiks,ref:babar_phik,ref:babar_etapks},
whereas they are predicted to be equal to within a few percent 
in the Standard Model~\cite{ref:smpollution}.
Since the former decays are predominantly governed
by flavour changing neutral current penguin amplitudes,
they are commonly believed to be sensitive to new physics contributions~\cite{ref:newphys}.

The significance of the deviation from the Standard Model prediction
will be clarified once more data has been accumulated and the 
statistical precision of the measurements improved.
However, it is essential to try to find additional modes 
which may be sensitive to the same effect, 
in order to corroborate the evidence and/or investigate which 
underlying interaction is responsible.
In this spirit, experimental results using the decay modes 
\mbox{$B^0 \to \etap \ks$},  
\mbox{$B^0 \to K^+ K^- \ks$} and 
\mbox{$B^0 \to \ks \pi^0$}
have already been produced~\cite{ref:babar_kspi0}, 
alongside those for $B^0 \to \phi \ks$.
Additional modes sensitive to the $b \to s$ penguin amplitude
would make extremely welcome additions to this set of measurements.

It is not only the $b \to s$ penguin transition which is of interest.
In general any decay mode which is sensitive to $CP$ violation,
which to date has only been observed in a handful of decay modes, 
is certainly worth pursuing.
Furthermore, any model which proposes new physics effects in 
$b \to s$ amplitudes via the introduction of a new heavy particle in the loop,
may also have an effect on the phase of other ``loop'' amplitudes, 
such as the $b \to d$ penguin or $B - \bar{B}$ mixing.

So, we are interested in any $B$ meson decay mode in which $CP$ violation,
and/or new physics contributions may be investigated.  
As has been known for some time, neutral $B$ decays to $CP$ eigenstates 
can exhibit $CP$ violation with rather straightforward 
phenomenology~\cite{ref:bigicartersanda}.
We are also interested in non-$CP$ eigenstate final states,
where new physics contributions have a relatively clean signal,
such as radiative $B$ decays.
In this article we present some examples of final states which 
we hope may provide fertile ground for investigating these effects
at current and future $B$ factories.

\section{\boldmath{$CP$} of \boldmath{$B^0 \to P^0P^0X^0$} Decays}

\subsection{Spin-0 \boldmath{$X^0$}}

We are considering decays of the form $B^0 \to P^0P^0X^0$.
Here $B^0$ represents a neutral $B$ meson 
(which can actually be either $B^0$ or $\bar{B}^0$);
$P^0$ represents any spin-0 neutral particle
(we use $P^0$ since this particle will typically be a pseudo-scalar 
($J^P = 0^-$), but scalar particles ($J^P = 0^+$) are also allowed).
$X^0$ also represents any spin-0 neutral particle,
which may be either identical to or different to $P^0$.
In what follows, $L$ denotes the angular momentum of the $P^0P^0$ system,
and $L^{\prime}$ denotes the angular momentum of 
$X^0$ relative to the $P^0P^0$ system.

Due to Bose-Einstein statistics, the $P^0P^0$ wave-function must be symmetric, 
and hence the angular momentum between the two $P^0$ particles ($L$) 
must be even.
Therefore, we can write down the $CP$ of the $P^0P^0$ system:
\begin{eqnarray}
  \label{eq:cp_p0p0}
  CP\left(P^0P^0\right) & = &
  C\left(P^0P^0\right) \times P\left(P^0P^0\right) =
  C\left(P^0\right)^2 \times P\left(P^0\right)^2 \times \left(-1\right)^{L}
  \\
  & = & CP\left(P^0\right)^2 = +1,
\end{eqnarray}
where we have assumed that $P^0$ is a $CP$ eigenstate 
({\it i.e.} $CP\left(P^0\right) = \pm 1$) in the last step.

By conservation of angular momentum in the decay $B^0 \to P^0P^0X^0$, 
we obtain
\begin{eqnarray}
  \label{eq:consJ} 
  {\bf J}_{B^0} & = &
  {\bf L} + {\bf L}^{\prime} + 
  {\bf S}_{P^0} + {\bf S}_{P^0} + {\bf S}_{X^0} 
  \\
  {\bf 0} & = & {\bf L} + {\bf L}^{\prime},
\end{eqnarray}
since the neutral $B$ meson is a spin-0 particle,
as are $P^0$ and $X^0$.
In the above equations, ${\bf J}$, ${\bf L}$ and ${\bf S}$ represent
the total, orbital and intrinsic angular momentum, respectively
(and elsewhere $J$, $L$ and $S$ represent their magnitudes).
Therefore, the angular momentum between the $P^0P^0$ system and $X^0$ 
($L^{\prime}$) must be equal to $L$, 
and we can write down the $CP$ of the $P^0P^0X^0$ system:
\begin{eqnarray}
  \label{eq:cp_p0p0x0}
  CP\left(P^0P^0X^0\right) & = & 
  CP\left(P^0P^0\right) \times CP\left(X^0\right) \times \left(-1\right)^{L^{\prime}}
  \\
  & = & CP\left(X^0\right).
\end{eqnarray}

Therefore, provided that both $P^0$ and $X^0$ are $CP$ eigenstates,
the $P^0P^0X^0$ system will also be a $CP$ eigenstate,
with the same $CP$ as the $X^0$.
The possible $P^0$ are
$\pi^0$, $\eta$, $\etap$, $f_0$, $a_0$, $\ks$, $\kl$, and $D_{CP}$; 
the possible $X^0$ include all the $P^0$ candidates, $\eta_c$ and $\chi_{c0}$.
Note that although $\ks$ and $\kl$ are not actually $CP$ eigenstates,
the $CP$ violation effect in the neutral kaon system 
is small enough to be ignored.
Furthermore, since $\ks$ would be reconstructed via its decay to 
$\pi^+\pi^-$ (or possibly $\pi^0\pi^0$), 
the final state is actually $\left( \pi\pi \right)_{K}$,
which {\it is} a $CP$ eigenstate.
Using this rationale we also consider $D_{CP}$ 
(a neutral $D$ meson reconstructed in a $CP$ eigenstate)
as a possible $P^0$; 
for completeness, $K^{*0}_0 (1430)$ is also a $P^0$ candidate 
when reconstructed in a $CP$ eigenstate ($K^{*0}_0 \to \ks \pi^0$).

We conclude that for any $CP$ eigenstate spin-0 neutral particles
$P^0$ and $X^0$ which might be reconstructed at a $B$ factory, 
the final state in $B^0 \to P^0P^0X^0$ decays is a $CP$ eigenstate,
and that these modes therefore provide numerous possibilities for
the study of time-dependent $CP$ violation.

\subsection{\boldmath{$X^0$} with Non-Zero Intrinsic Spin}

In the case that $X^0$ has non-zero angular momentum,
the criterion to conserve angular momentum (Eq.~\ref{eq:consJ}) becomes
\begin{equation}
  {\bf 0} = {\bf L} + {\bf L}^{\prime} + {\bf S}_{X^0}.
\end{equation}

The $P^0P^0$ system is required to have even angular momentum, 
as noted above.
In the case that $L = 0$, we find 
$L^{\prime} = S_{X^0}$,
and the final state has definite $CP$ content.
However, we must also include the possibility that 
$L = 2,4,\ldots$,
where the final state contains both $CP$ even and $CP$ odd components.
Although states with large values of $L$ are suppressed due to 
the angular momentum barrier, some experimental study is required in order to
determine the $CP$ composition in these final states.
In the case that the spin-0 state of $P^0P^0$ is dominant, 
these modes will be particularly worthy of further study.

\subsection{\boldmath{$X^0 = \gamma$}}

We also consider the special case where $X^0 = \gamma$.
Since the photon is a massless vector particle, 
${\rm spin}\left( P^0 P^0 \right) = 0$ is forbidden.
Therefore, spin-2 is the lowest spin state possible for the $P^0P^0$ system.

Decays of the form $B^0 \to M^0\gamma$ are of considerable 
theoretical interest~\cite{ref:atwood_gronau_soni}.
Here $M^0$ can be any hadronic self-conjugate $CP$ eigenstate,
and $M^0\gamma$ is not required to be a $CP$ eigenstate since this does
not affect the time-dependent asymmetry.

Since the $P^0P^0$ system must contain an even number (0, 2 or 4) of each 
quark + anti-quark flavour, decays of the form 
$B^0 \to P^0 P^0 \gamma$ can only occur via $b \to d \gamma$ transitions.
Other decay modes sensitive to the same amplitude 
(typically $B^0 \to \rho^0 \gamma$, $B^0 \to \omega \gamma$)
suffer from considerable experimental difficulties. 
For that reason, modes such as $B^0 \to \eta \eta \gamma$, 
$B^0 \to \etap \etap \gamma$ and 
$B^0 \to \ks \ks \gamma$ may be useful to study mixing-induced 
asymmetries in $b \to d \gamma$ decays.

Since the $P^0P^0$ system must have spin of at least two,
decays of this form may be suppressed due to the angular momentum barrier.
However, evidence for a radiative \mbox{$b \to s$} transition
resulting in a tensor meson, 
has been observed with a branching fraction of 
${\cal B}\left( B^0 \to K_2^{*0}(1430) \gamma \right) = 
{\cal O}\left( 10^{-5} \right)$~\cite{ref:kpipigamma}.
Thus, observation of similar decays mediated by the $b \to d$ transition
({\it e.g.} 
$B \to f_2(1270) \gamma, a_2(1320) \gamma, f_2^{\prime}(1525) \gamma$) 
via $P^0P^0\gamma$ final states may soon be within reach of the $B$ factories.

\subsection{\boldmath{$CP$} of \boldmath{$B^0 \to P^0Q^0X^0$} Decays}

It may be noted that the use of Bose-Einstein statistics in the 
above argument is, in fact, not necessary.
If we consider decays of the type $B^0 \to P^0Q^0X^0$~\cite{ref:dqstl},
where the three final state particles are different 
$CP$ eigenstate spin-0 neutral particles, Eq.~\ref{eq:consJ} becomes
\begin{equation}
  {\bf 0} = {\bf L} + {\bf L}^{\prime} + 
  {\bf S}_{P^0} + {\bf S}_{Q^0} + {\bf S}_{X^0},
\end{equation}
and so we have $L = L^{\prime}$ as before. 
(Here ${\bf L}$ denotes the angular momentum between $P^0$ and $Q^0$, 
and ${\bf L^{\prime}}$ denotes the angular momentum of $X^0$ 
relative to the $P^0-Q^0$ system.)
We can then write down the $CP$ of the final state,
\begin{eqnarray}
  \label{eq:cp_p0q0x0}
  CP\left(P^0Q^0X^0\right) & = & 
  CP\left(P^0\right) \times CP\left(Q^0\right) \times CP\left(X^0\right) 
  \times \left(-1\right)^{L} \times \left(-1\right)^{L^{\prime}} 
  \\
  & = & 
  CP\left(P^0\right) \times CP\left(Q^0\right) \times CP\left(X^0\right).
\end{eqnarray}
  
This realisation greatly increases the number of possible final states. 
However, in many cases, decays with three different final state
particles may have contributions from a number of different amplitudes,
complicating the phenomenological interpretation of their time-dependence.
Exceptions may arise in the case that two or more of the particles are 
from the set $\left( \pi^0, \eta, \etap \right)$,
or when the final state includes $\ks$ and $\kl$;
nevertheless, we restrict ourselves to the case $P^0 = Q^0$ 
in the main part of this paper.
Note however one useful consequence of this observation:
in the case of studying time-dependent $CP$ violation in a decay
such as $B^0 \to \etap K^{*0}$ with $K^{*0} \to \ks \pi^0$,
any background from $B^0 \to \etap \ks \pi^0$,
either non-resonant or produced via some other intermediate state,
can be treated as signal since it has the same $CP$ eigenstate.
(Possible contributions from different amplitudes may however
pollute the interpretation of the result.)

\section{Consideration of Specific \boldmath{$B^0 \to P^0P^0X^0$}}

In the remainder of this article, we consider specific decays of the form 
$B^0 \to P^0P^0X^0$ where both $P^0$ and $X^0$ are spin-0 neutral particles. 
As is clear from Table~\ref{tab:p0p0x0}, 
there are an enormous number of modes of this type.
We should consider which of these are most likely to be useful,
in terms of both a clean phenomenological interpretation,
and of the possibility of obtaining a sufficiently large sample
to study $CP$ violating effects.
This last point is problematic, 
since most of these decay modes have not yet been observed,
and we know of no reliable technique with which to estimate
the three body branching fractions.
Where necessary, we guess the approximate size of the branching fractions
based on existing experimental measurements.
As a corollary, we note that measurements of the branching fractions 
of these modes
will provide useful information on the nature of hadronic $B$ decays.

\begin{center}
  \begin{table}
\begin{tabular}{|c|cccccccc|}
  \hline
  & \multicolumn{8}{c|}{$P^0$} \\
  \hline
  $X^0$ & $\pi^0$ & $\eta$ & $\etap$ & $\fz$ & $\az$ & $\ks$ & $\kl$ & $\dcp$ \\
  \hline
  $\pi^0$ & \ul{$\pi^0\pi^0\pi^0$} & $\eta\eta\pi^0$ & $\etap\etap\pi^0$ & $\fz\fz\pi^0$ & $\az\az\pi^0$ & \ul{$\ks\ks\pi^0$} & $\kl\kl\pi^0$ & $\dcp\dcp\pi^0$ \\
  $\eta$  & $\pi^0\pi^0\eta$  & $\eta\eta\eta$  & $\etap\etap\eta$  & $\fz\fz\eta$  & $\az\az\eta$  & $\ks\ks\eta$  & $\kl\kl\eta$  & $\dcp\dcp\eta$  \\
  $\etap$ & $\pi^0\pi^0\etap$ & $\eta\eta\etap$ & $\etap\etap\etap$ & $\fz\fz\etap$ & $\az\az\etap$ & $\ks\ks\etap$ & $\kl\kl\etap$ & $\dcp\dcp\etap$ \\
  $\fz$   & $\pi^0\pi^0\fz$   & $\eta\eta\fz$   & $\etap\etap\fz$   & $\fz\fz\fz$   & $\az\az\fz$   & $\ks\ks\fz$   & $\kl\kl\fz$   & $\dcp\dcp\fz$   \\
  $\az$   & $\pi^0\pi^0\az$   & $\eta\eta\az$   & $\etap\etap\az$   & $\fz\fz\az$   & $\az\az\az$   & $\ks\ks\az$   & $\kl\kl\az$   & $\dcp\dcp\az$   \\
  $\ks$   & \ul{$\pi^0\pi^0\ks$}   & \ul{$\eta\eta\ks$}   & \ul{$\etap\etap\ks$}   & $\fz\fz\ks$   & $\az\az\ks$   & \dul{$\ks\ks\ks$}   & $\kl\kl\ks$   & $\dcp\dcp\ks$   \\
  $\kl$   & $\pi^0\pi^0\kl$   & $\eta\eta\kl$   & $\etap\etap\kl$   & $\fz\fz\kl$   & $\az\az\kl$   & \ul{$\ks\ks\kl$}   & $\kl\kl\kl$   & $\dcp\dcp\kl$   \\
  $\dcp$  & $\pi^0\pi^0\dcp$  & $\eta\eta\dcp$  & $\etap\etap\dcp$  & $\fz\fz\dcp$  & $\az\az\dcp$  & \ul{$\ks\ks\dcp$}  & $\kl\kl\dcp$  &  \\
  $\etac$ & $\pi^0\pi^0\etac$ & $\eta\eta\etac$ & $\etap\etap\etac$ & $\fz\fz\etac$ & $\az\az\etac$ & $\ks\ks\etac$ & $\kl\kl\etac$ &  \\
  $\chic$ & $\pi^0\pi^0\chic$ & $\eta\eta\chic$ & $\etap\etap\chic$ & $\fz\fz\chic$ & $\az\az\chic$ & $\ks\ks\chic$ & $\kl\kl\chic$ &  \\
  \hline
\end{tabular}
\caption{
  \label{tab:p0p0x0}
  Possible $B^0 \to P^0P^0X^0$ final states.
  Underlined modes are discussed in detail in the text.
  The doubly underlined mode $B^0 \to \ks \ks \ks$ has already been observed.
}
  \end{table}
\end{center}

\subsection{\boldmath{$X^0 = \eta_c, \chi_{c0}$}}

\begin{figure}[!tb]
  \begin{center} 
    \includegraphics[width=0.5\textwidth]{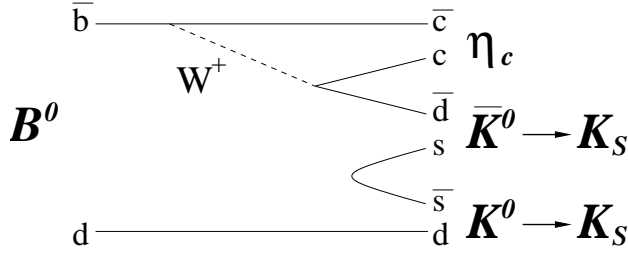}   
  \end{center}     
  \caption{       
    \label{fig:etacksks}       
    Feynman diagram for $B^0 \to \eta_c \ks \ks$,         
    which is the same as that for $B^0 \to \eta_c \pi^0$           
    with additional $s\bar{s}$ production.
    Only the (dominant) tree contribution is shown.
  }               
\end{figure}

In the study of $B \to {\rm charmonium}$ decays,
$c\bar{c} = J/\psi$ is usually preferred over 
$c\bar{c} = \eta_c \ {\rm or} \ \chi_{c0}$,
owing to the former's clean signal and high reconstruction efficiency
in the decay $J/\psi \to l^+l^-$, $l = e,\mu$.
Although in principle $B^0 \to \eta_c \ks \ks$ or $B^0 \to \eta_c \pi^0\pi^0$, 
and equivalent modes for $\chi_{c0}$,
could be used to investigate the $b \to c \bar{c} d$ transition, 
as shown in Fig.~\ref{fig:etacksks},
in practise the backgrounds will
be rather large and the reconstruction efficiency rather low.
We do not consider these modes further.

\subsection{\boldmath{$P^0 = D_{CP}$} or \boldmath{$X^0 = D_{CP}$}}

Decays of the form $B \to D D K$ have previously been shown to be useful
in probing the $b \to c \bar{c} s$ amplitude and in resolving the 
ambiguity in $\phi_1$ from measurements of $\sinbb$~\cite{ref:ddk}.
There is no immediately apparent benefit from requiring the $D$ meson
to be reconstructed in a $CP$ eigenstate.
The same applies for $B \to D D \pi$, which probes $b \to c \bar{c} d$.

\begin{figure}[!tb]
  \begin{center} 
  \includegraphics[width=\textwidth]{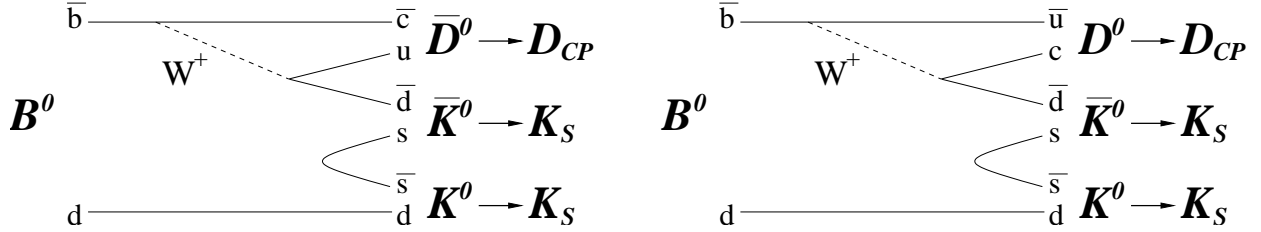}
  \end{center}       
\caption{         
  \label{fig:dcpksks}  
  Feynman diagrams for $B^0 \to D_{CP} \ks \ks$,
  which are the same as those for $B^0 \to D_{CP} \pi^0$
  with additional $s\bar{s}$ production.
}                 
\end{figure}

The decay mode $B^0 \to D_{CP} \pi^0$ has been proposed to determine
the $B - \bar{B}$ mixing phase with an order of magnitude less
theoretical uncertainty than exists in $B^0 \to J/\psi \ks$~\cite{ref:dcppi0}.
Due to this theoretical cleanliness, this mode is also sensitive 
to some new physics models~\cite{ref:newphys}.
The decay $B^0 \to D_{CP} \ks \ks$ is mediated by the same diagrams, 
with additional $s \bar{s}$ production, 
as shown in Fig.~\ref{fig:dcpksks}.

At $B$ factories, the displaced vertex and the narrow width
of the $K_S$ meson provide a rather clean experimental signature.
This leads to lower backgrounds from random combinations of particles
than in corresponding modes containing $\pi^0$s.
Therefore, although $B^0 \to D_{CP} \ks \ks$ is expected to have a
smaller branching fraction than $B^0 \to D_{CP} \pi^0$,
the sensitivity to time-dependent observables may be comparable.

Similar decay modes $B \to DKK$ have previously been observed,
with branching fractions of the order of $10^{-4}$~\cite{belle_dkk}.
The production may be dominated by intermediate spin-1 $KK$ resonances,
but a significant non-resonant contribution is not ruled out.
In this case, it may be possible to observe 
the decay $B^0 \to \bar{D}^0 \ks \ks$, with the current $B$ factory statistics,
using the decay $\bar{D}^0 \to K^+\pi^-$.
As with $B^0 \to D_{CP} \pi^0$, the requirement to reconstruct the $D$ meson
in a $CP$ eigenstate leads to a further reduction in statistics;
however it may be possible to perform time-dependent measurements of
$B^0 \to D_{CP} \ks \ks$ in the future.

\subsection{\boldmath{$P^0$ and $X^0 = \ks, \kl$}}

\begin{figure}[!tb]
  \begin{center}
    \includegraphics[width=12cm]{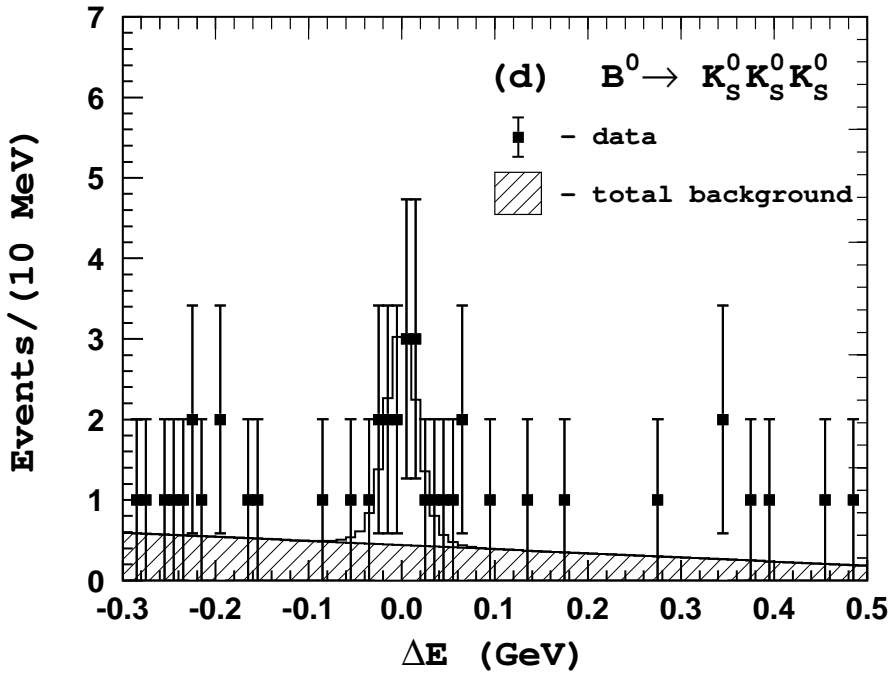}
  \end{center}
  \caption{
    \label{fig:ksksks}
    Observation of $B^0 \to \ks \ks \ks$, from~\cite{ref:kkk}.
    The plotted variable, $\Delta E$ is the difference between
    the reconstructed and expected $B$ candidate energies.
    From $78 \ {\rm fb}^{-1}$ of data recorded on the $\Upsilon(4S)$ resonance,
    a signal yield of $12.2^{+4.5}_{-3.8}$ is obtained, 
    leading to a branching fraction of 
    ${\cal B}\left(B^0 \to \ks \ks \ks\right) = 
    \left( 4.2^{+1.6}_{-1.3} \pm 0.8 \right) \times 10^{-6}$.
  }
\end{figure}

As previously mentioned, 
the possibility of a new physics signal in the time-dependence of 
$B^0 \to \phi \ks$ decays~\cite{ref:belle_phiks} 
is one of the most exciting results in high energy physics today.
To fully investigate this possibility, 
as many $b \to s\bar{s}s$ modes as possible need to be utilized.
In addition to $\phi \ks$, $\etap \ks$ and $K^+K^- \ks$ have been
used to date.
In the latter case, an isospin analysis indicates that after removing events 
where the $K^+K^-$ combination may have originated from a $\phi$ meson, 
the remaining $K^+K^-\ks$ candidates are predominantly $CP = +1$, 
which in turn indicates that the $K^+K^-$ system has even angular 
momentum~\cite{ref:kkk}.
Since any even angular momentum state which decays to $K^+K^-$
should also decay to $\ks \ks$ and $\kl \kl$, this further suggests that 
$B^0 \to \ks \ks \ks$ and $B^0 \to \kl \left(\kl \ks\right)_{{\rm non}-\phi}$ 
may have reasonable branching fractions.
Indeed, $B^0 \to \ks \ks \ks$ has been observed,
as shown in Fig.~\ref{fig:ksksks}.

In terms of phenomenology, 
these decays have an advantage over $B^0 \to K^+K^- \ks$, 
since the latter suffers a contribution from the $b \to u$ tree diagram, 
which has a different weak phase.
Here, however, there is no $u$ quark in the final state.
The $b \to s u \bar{u}$ tree diagram
followed by rescattering into $s d \bar{d}$ or $s s \bar{s}$ 
is OZI-suppressed.
Therefore these are almost pure penguin decays.
There can be contributions from both 
$b \to s s \bar{s}$ with additional $d \bar{d}$ production, and 
$b \to s d \bar{d}$ with additional $s \bar{s}$ production, 
but these diagrams have the same weak phase.
Any new physics contribution expected in $B^0 \to \phi \ks$ also affects
$B^0 \to \ks\ks\ks$ and $B^0 \to \kl \left(\kl \ks\right)_{{\rm non}-\phi}$
(and, for completeness, $B^0 \to \phi \kl$)
and in the absence of new physics all should exhibit the same $CP$ violating
effects as $J/\psi \ks$.

Turning to experimental considerations,
we note that final states containing $\kl$ tend to suffer large backgrounds.
This appears to be an insurmountable problem for $B$ decays containing 
two or more $\kl$ mesons in the final state at any planned experiment,
and so we do not consider $B^0 \to \ks \kl \kl$ or $B^0 \to \kl \kl \kl$
further.

The first results on time-dependent analyses from the $B$ factories
have used $B$ decays with final states including particles which 
may be considered to originate from the $B$ decay vertex,
{\it e.g.} $J/\psi \ks$, $\pi^+\pi^-$, $\phi \ks$.
For these modes, the performances of the vertexing subsystems and algorithms
are well understood~\cite{ref:belle_sin2phi1,ref:babar_sin2phi1,ref:resolution_nim}.
Results using modes where the 
primary $B$ daughters have a lifetime which is not negligible,
{\it e.g.} $D^{*+}D^{-}$, $D^{*+}D^{*-}$, 
have also appeared~\cite{ref:babar_dd}.
In these analyses, the displacement of the daughter tracks 
must be taken into account when measuring the vertex position.
Recently, the BaBar collaboration have succeeded to measure 
the $B$ decay vertex in $B^0 \to \ks \pi^0$~\cite{ref:babar_kspi0}.
The possibility to obtain vertex information from $\ks$ mesons 
makes time-dependent analyses of 
$B^0 \to \ks\ks\ks$ and $B^0 \to \ks\ks\kl$ feasible.

\subsubsection{\boldmath{$B^0 \to \ks \ks \ks$}}

As pointed out above, this mode has been observed~\cite{ref:kkk}.
From $78 \ {\rm fb}^{-1}$ of data recorded on the $\Upsilon(4S)$ resonance, 
$12.2^{+4.5}_{-3.8}$ signal events are found.
From Fig.~\ref{fig:ksksks}, we can count the number of candidates
in the region $-0.1 \ { \rm GeV} < \Delta E < 0.1 \ { \rm GeV}$, 
to estimate the signal-to-background ratio ($S/B$).  
There are 21 candidates in this region giving $S/B \gtrsim 1$.

The efficiency to obtain a $\ks$ vertex reflects the
probability for the $\ks$ to decay inside the vertex detector,
and so depends on the $\ks$ momentum and on the size of the vertex detector.
In a three body $B^0 \to \ks\ks\ks$ decay, 
at least one $\ks$ must have fairly low momentum in the $B^0$ rest frame.
Therefore, we expect a high vertex efficiency for $B^0 \to \ks\ks\ks$,
if the vertex efficiency for $B^0 \to \ks \pi^0$ 
(where the $\ks$ has high momentum) is moderate.
In the time-dependent analysis of $B^0 \to \ks \pi^0$ 
by the BaBar collaboration, 
a vertex efficiency of 65\% is obtained~\cite{ref:babar_kspi0}.
Therefore, we expect that vertex efficiencies for $B^0 \to \ks\ks\ks$ 
of close to 100\% may be obtained at $B$ factories with similar
vertex detector geometry.

Noting that the purity of the $\ks \ks \ks$ signal is comparable
to that of $\etap \ks$, the precision on the parameter $S_{\ks \ks \ks}$
can be predicted~\cite{ref:superb_loi}.
From $140 \ {\rm fb}^{-1}$ Belle has 421 $B \to \etap \ks$ candidates,
and measures the error on $S_{\etap \ks}$ to be $\delta S_{\etap \ks} = 0.27$,
hence the statistical sensitivity from $B^0 \to \ks \ks \ks$ 
with $5 \ {\rm ab}^{-1}$ of data is estimated as $\delta S_{\ks \ks \ks} \sim 0.14$.

In fact, an initial time-dependent study of this mode may be practicable
at present $B$ factories as a number of improvements are possible:
\begin{enumerate}
\item By optimizing event selection criteria in each flavor tagging category.
  Events with good flavor tagging quality, which contribute most to the 
  statistical significance of the result, tend to have smaller 
  backgrounds from continuum processes ($e^+e^- \to q\bar{q}$, $q = u,d,s,c$).
\item By optimizing the $\ks$ selection for this analysis,
  some improvement in the $\ks$ efficiency may be obtained.
  The $\ks \ks \ks$ efficiency will benefit by the improvement to
  the third power.
\item By including events where one $\ks$ is reconstructed in the 
  $\pi^0\pi^0$ final state. Since 
  ${\cal B}\left( \ks \to \pi^+\pi^- \right) \approx 2 
  {\cal B}\left( \ks \to \pi^0\pi^0 \right)$, we expect
  \begin{equation}
    N_{\rm true} 
    \left( 
      B^0 \to 2 \left( \pi^+\pi^- \right)_{\ks} \left( \pi^0\pi^0 \right)_{\ks}
    \right) 
    \approx \frac{3}{2}
    N_{\rm true}
    \left( 
      B^0 \to 3 \left( \pi^+\pi^- \right)_{\ks}
    \right),
  \end{equation}
  where $N_{\rm true}$ indicates the number of $B$ mesons decaying in the 
  specified manner, 
  {\it i.e.} no reconstruction efficiency is taken into account.

  To estimate the ratio of reconstruction efficiencies,
  $\epsilon_{\ks \to \pi^+\pi^-}/\epsilon_{\ks \to \pi^0\pi^0}$,
  we refer to the most recent measurement of $\sinbb$~\cite{ref:sin2phi1_140fb}
  by the Belle collaboration, in which 
  1997 $B^0 \to J/\psi \left( \pi^+\pi^- \right)_{\ks}$ candidates and
  288 $B^0 \to J/\psi \left( \pi^0\pi^0 \right)_{\ks}$ candidates 
  are reconstructed. Taking background fractions into account
  \begin{equation}
    \frac{\epsilon_{\ks \to \pi^+\pi^-}}{\epsilon_{\ks \to \pi^0\pi^0}} 
    \approx
    \frac{1}{2}\frac{1997 \times 0.976 }{288 \times 0.82} \approx 4.1,
  \end{equation}
  and so we expect to find the reconstructed numbers of events 
  ($N_{\rm rec}$) to be approximately related by
  \begin{equation}
    N_{\rm rec} 
    \left( 
      B^0 \to 2 \left( \pi^+\pi^- \right)_{\ks} \left( \pi^0\pi^0 \right)_{\ks}
    \right) 
    \approx 
    0.36 \,
    N_{\rm rec}
    \left( 
      B^0 \to 3 \left( \pi^+\pi^- \right)_{\ks}
    \right).
  \end{equation}
  Although when one $\ks$ is reconstructed via $\pi^0\pi^0$, 
  both $S/B$ and $\epsilon_{\rm vtx}$ will be degraded,
  including these events will nonetheless help to reduce the statistical error.
\end{enumerate}

Although most of the background for this decay is expected to originate
from continuum processes, care will have to be taken to reject background
from $b \to c$ transitions.
Whilst $\eta_c \to \ks \ks$ and $J/\psi \to \ks \ks$ are forbidden
due to parity and $CP$ conservation respectively,
some other charmonium states,
({\it e.g.} $B^0 \to \left( \ks \ks \right)_{\chi_{c0}} \ks$),
may cause a background.
Such contributions can be measured from amplitude (Dalitz plot)
analyses of $B^0 \to \ks \ks \ks$ and other $B \to KKK$ decays,
or estimated from known branching fractions.
For example, if we assume that
${\cal B} \left( B^0 \to \chi_{c0} \ks \right) = 
\frac{1}{2} {\cal B} \left( B^+ \to \chi_{c0} K^+ \right) = 
\frac{1}{2} \left( 6.0 ^{+2.1}_{-1.8} \pm 1.1 \right) \times 10^{-4}$~\cite{ref:chic0k}, 
where the factor of $\frac{1}{2}$ is due to 
${\cal B}\left( K^0 \to \ks \right)$,
and take
${\cal B} \left( \chi_{c0} \to \ks \ks \right) = 
\left( 2.0 \pm 0.6 \right) \times 10^{-3}$~\cite{ref:pdg},
we estimate the background level,
$B^0 \to \left( \ks \ks \right)_{\chi_{c0}} \ks \sim 
0.1 \times {\cal B}\left(B^0 \to \ks \ks \ks\right)$.
Backgrounds of this type can be rejected using invariant mass requirements.

To summarize this subsection, $B^0 \to \ks \ks \ks$ is 
(a) theoretically clean, 
(b) very interesting, 
(c) already observed, 
(d) an analysis which requires a great deal of further work!

\subsubsection{\boldmath{$B^0 \to \ks \ks \kl$}}

We can compare $B^0 \to \ks \ks \kl$ to $B^0 \to \ks \ks \ks$ 
again by using information from the most recent measurement of 
$\sinbb$~\cite{ref:sin2phi1_140fb}
by the Belle collaboration,
and comparing the numbers of candidates of
$J/\psi \left( \pi^+\pi^- \right)_{\ks}$ and $J/\psi \kl$, 
taking background fractions into account:
\begin{equation}
  \epsilon_{\ks \to \pi^+\pi^-}/\epsilon_{\kl} \approx
  \frac{1997 \times 0.976 }{2332 \times 0.60} \approx 1.4.
\end{equation}
Therefore, assuming equal branching fractions, we might expect almost as many 
$B^0 \to \ks \ks \kl$ as $B^0 \to \ks \ks \ks$ events,
albeit with worse $S/B$ and $\epsilon_{\rm vtx}$.
As above, there may be some gain by including events where one 
$\ks$ is reconstructed via $\pi^0\pi^0$.
To know the feasibility of time-dependent $CP$ analyses using 
$\ks \ks \kl$
requires more detailed knowledge of the size of the background.

Since this final state has the opposite $CP$ to $\ks \ks \ks$,
there is a significant benefit (not least in the presentation of results)
in studying these two modes in conjunction with each other.
However, there is an additional constraint which may be used in this case.
Recalling that the measurement of the $CP$ composition of 
$B^0 \to \left( K^+K^- \right)_{{\rm non}-\phi} \ks$ as predominantly
$CP$ even suggests a reasonably large branching fraction for
$B^0 \to \left( \ks \ks \right)_{{\rm non}-\phi} \ks$;
by the same argument a rather small branching fraction for
$B^0 \to \left( \ks \kl \right)_{{\rm non}-\phi} \ks$ is expected.
Since $\phi \to \ks \kl$ is known to have a large branching fraction, 
we therefore have nothing to lose by imposing a $\phi$ mass constraint.
Since the $\phi$ resonance is rather narrow, 
this constraint may be extremely useful in rejecting background.
Further experimental investigation is required.

\subsection{\boldmath{$P^0 = \pi^0, \eta, \etap$ and $X^0 = \ks, \kl$}}

The decay $B^+ \to \phi \phi K^+$ has previously been pointed out to 
be sensitive to possible new physics contributions~\cite{ref:hazumi_phiphik},
and has recently been observed~\cite{ref:phiphik}.
Similarly, $B^0 \to P^0 P^0 \ks$ modes where $P^0$ has an $s\bar{s}$ component,
may also be sensitive to new physics.
As the branching fractions for the decays $B \to \etap K$
and $B \to \etap X_S$ are known to be larger than 
theoretical expectations, $B \to \etap \etap K$ decays
appear particularly interesting.
The dominant amplitude for these decays is the $b \to s$ penguin,
although other contributions are possible.
Thus, studies of the time-dependence of $B^0 \to \etap \etap \ks$
can also help to probe for new physics phases in these amplitudes.

As mentioned above, there has recently been some activity on 
time-dependent analysis of $B^0 \to \ks\pi^0$~\cite{ref:babar_kspi0}, 
which has contributions
from the $b \to s \bar{d} d$ penguin and $b \to s \bar{u} u$
tree and penguin amplitudes.
By adding a $u\bar{u}$ or $d\bar{d}$ pair as appropriate, 
the same processes can result in the $\ks \pi^0\pi^0$ final state.
In the three body decay mode the vertexing efficiency should be 
improved compared to $\ks \pi^0$ since the $\ks$ will have lower momentum.
(Assuming a phase space distribution of the $\ks \pi^0\pi^0$ decay products,
we expect the vertexing inefficiency to be reduced by $10-20\%$
compared to $\ks \pi^0$.)

Measured branching fractions for $B^+ \to K^+ \pi^+ \pi^-$
and $B^0 \to \ks \pi^+ \pi^-$~\cite{ref:kkk} suggest that 
$B^0 \to \ks \pi^0 \pi^0$ may have a larger branching fraction 
than $B^0 \to \ks\pi^0$.
Furthermore,
an amplitude analysis of $B^+ \to K^+ \pi^+ \pi^-$ has shown 
that there are large contributions from $B^+ \to K^*_0(1430)^0 \pi^+$,
$B^+ \to f_0(980)K^+$ and a non-resonant source~\cite{ref:belle-conf-0338}.
In each case, performing an isospin rotation on the spectator quark
leads to an amplitude which contributes to the $B^0 \to \ks \pi^0\pi^0$
final state.
Therefore, we expect that the branching fraction for 
$B^0 \to K^0 \pi^0\pi^0$ should be comparable to those for
$B^0 \to K^0 \pi^+ \pi^-$ and $B^+ \to K^+ \pi^+ \pi^-$
(which are $\sim 5 \times 10^{-5}$).

Large backgrounds would be expected in this mode,
but these appear to be experimentally controllable.
Continuum background can be handled in the usual manner,
contributions which proceed via charmed intermediate states
({\it e.g.} $B^0 \to \bar{D}^0 \pi^0$ or $B^+ \to \bar{D}^0 \rho^+$ 
with $\bar{D}^0 \to K^0 \pi^0$)
can be removed using two-particle invariant mass requirements,
and other charmless $B$ decays ({\it e.g.} $B^0 \to \ks \pi^0$) 
are intrinsically small since their branching fractions are
less than that expected for $B^0 \to K^0 \pi^0\pi^0$.
Background from $B^0 \to K^{*0} \gamma$ with $K^{*0} \to K^0 \pi^0$
is also expected to be small, 
and can be further reduced making a requirement on the 
energy asymmetry of the clusters of the neutral pion candidates.

Thus $B^0 \to \ks \pi^0 \pi^0$
may provide another handle to probe the $b \to s$ transition.
Using the branching fraction given above,
and taking subdecays and efficiencies into account,
we expect the statistical errors of the time-dependent parameters in this mode 
will be comparable to those obtained from $B^0 \to \ks \pi^0$.
Thus we anticipate a precision of $\delta S_{\ks \pi^0\pi^0} \sim \delta S_{\ks \pi^0} \sim 0.10$
will be obtained with $5 \ {\rm ab}^{-1}$ of data~\cite{ref:superb_loi}.

\subsection{\boldmath{$P^0 = \ks, \kl$ and $X^0 = \pi^0, \eta, \etap$}}

Charmless $B$ decays to final states containing an even number of 
$s\bar{s}$ quarks are suppressed in the Standard Model.
These final states can be produced either via
a $b \to d$ penguin transition, or by $s\bar{s}$ production following 
a decay to a final state containing no strange quarks,
and both of these amplitudes are rather small. 
No such decay has been observed to date
(although evidence for $B^+ \to K^+ K^- \pi^+$ has been found~\cite{ref:kkk}).
Nonetheless, branching fraction estimates for some hadronic 
$b \to d$ transitions suggest these will soon be within reach of 
the $B$ factories, creating some interesting possibilities.
The decay $B^0 \to \ks \ks \pi^0$ may be mediated by the
$b \to d \bar{s} s$ penguin 
({\it e.g.} $B^0 \to \ks K^{*0}$ with $K^{*0} \to \ks \pi^0$),
or by the same diagrams which mediate $B^0 \to \pi^0 \pi^0$ followed
by $s\bar{s}$ production.
In either case, interesting observables may be investigated,
but some effort will be required to disentangle the contributing
amplitudes.

\subsection{\boldmath{$P^0$ and $X^0 = \pi^0, \eta, \etap$}}

Once again noting that final states containing even numbers of $s + \bar{s}$
quarks are suppressed, any combination of 
$P^0,X^0 = \left( \pi^0, \eta, \etap \right)$
should be produced via $u\bar{u}$ and $d\bar{d}$ components only.
Such two-body final states are used to determine $\phi_2$,
one of three angles of the Unitarity Triangle, 
and the three body final states also
contain useful information, in principle.
One interesting scenario would be modes which are produced by the 
same amplitudes as $B^0 \to \pi^0 \pi^0$, but for which it is
possible to reconstruct a vertex position, 
and hence study the time-dependence.
Additional $s\bar{s}$ production can result in $\ks \ks \pi^0$;
additional $d\bar{d}$ or $u\bar{u}$ production may result in
$\eta^{(\prime)} \eta^{(\prime)} \pi^0$.
The first evidence for $B^0 \to \pi^0 \pi^0$ has recently been 
seen~\cite{ref:pi0pi0}, with a branching fraction of around
${\cal B}\left( B \to \pi^0\pi^0 \right) \approx 2 \times 10^{-6}$.
This suggests that extremely large data samples would be necessary
in order to perform a time-dependent analysis on these final states.

The angle $\phi_2$ can also be determined from the Dalitz plot of
$B^0 \to \pi^+ \pi^- \pi^0$, 
or from an isospin analysis of $B \to \rho \pi$ decays~\cite{ref:snyder-quinn}.
It has been suggested that these 
analyses may be complicated by the presence of a broad scalar $\sigma$
resonance decaying to $\pi^+\pi^-$~\cite{ref:pipipi0}.  
The study of $B^0 \to \pi^0\pi^0\pi^0$ could limit the size of this effect,
since any such resonance must decay also to $\pi^0\pi^0$.
Therefore, it could limit the contribution from this class of background to 
the $B \to \rho\pi$ final states 
($\rho^+\pi^-$, $\rho^-\pi^+$, $\rho^+\pi^0$, $\rho^0\pi^+$, $\rho^0\pi^0$),
all of which have now been observed~\cite{ref:rhopi,ref:rho0pi0},
except for $B^0 \to \rho^0\pi^0$ for which evidence has been 
uncovered~\cite{ref:rho0pi0}.

\section{Summary}
It is clear that there are an enormous number of possible $B^0 \to P^0P^0X^0$
decay modes which could yield important information given enough 
integrated luminosity.  
Time-dependent studies of many of these modes appear 
experimentally feasible with the 
large data samples which can be obtained at a ``Super $B$ factory''.
For modes which use multiple photons in the reconstruction,
the environment at $e^+e^-$ machines is preferable to that at hadron colliders.
Furthermore, it is only at $e^+e^-$ machines that 
precise knowledge of the interaction point can be obtained to allow
the $B$ vertex to be reconstructed from a single $\ks$ meson.
Reconstruction of the $B$ vertex from multiple $\ks$ mesons 
({\it e.g.} in $B^0 \to \ks\ks\ks$) at hadron colliders,
appears experimentally challenging, but not impossible.

Of the possible $B^0 \to P^0P^0X^0$ decays,
$B^0 \to \ks \ks \ks$ appears to be an extremely promising mode
to study the $b \to s \bar{s} s$ transition and search for phases 
from new physics. 
First results with this mode may be obtained at
the present $B$ factories in the near future.
In addition, $B^0 \to D_{CP} \ks \ks$,
$B^0 \to \etap \etap \ks$, $B^0 \to \pi^0\pi^0 \ks$ and $B^0 \to \pi^0 \pi^0 \pi^0$
appear deserving of further study.

\section{Acknowledgements}

We are grateful to A.~Bondar, S.~Hashimoto, Y.~Okada, T.~Shibata and A.~Soni
for fruitful discussions.
T.~G. is supported by the Japan Society for the Promotion of Science.


\begin{thebibliography}{99}
\bibitem{ref:belle_sin2phi1}
  K.~Abe, {\it et al.} (Belle Collaboration),
  Phys. Rev. Lett. {\bf 87}, 091802 (2001);
  Phys. Rev. D {\bf 66}, 032007 (2002);
  Phys. Rev. D {\bf 66}, 071102 (2002).
\bibitem{ref:babar_sin2phi1}
  B.~Aubert {\it et al.} (BaBar Collaboration),
  Phys. Rev. Lett. {\bf 87}, 091801 (2001);
  Phys. Rev. D {\bf 66}, 032003 (2002);
  Phys. Rev. Lett {\bf 89}, 201802 (2002).  
\bibitem{ref:eoe}
  J.~R.~Batley {\it et al.} (NA48 Collaboration),
  Phys. Lett. B {\bf 544}, 97 (2002);
  A.~Alavi-Harati {\it et al.} (KTeV Collaboration),
  Phys. Rev. Lett. {\bf 83}, 22 (1999),
  Phys. Rev. D {\bf 67}, 012005 (2003).
\bibitem{ref:km}
  M.~Kobayashi and T.~Maskawa, 
  Prog. Theor. Phys. {\bf 49}, 652 (1973).
\bibitem{ref:belle_phiks}
  K.~Abe {\it et al.} (Belle Collaboration), 
  Phys. Rev. D {\bf 67}, 031102 (2003);
  K.~Abe {\it et al.} (Belle Collaboration), 
  Phys. Rev. Lett. {\bf 91}, 261602 (2003).
\bibitem{ref:babar_phik}
  B.~Aubert {\it et al.} (BaBar Collaboration),
  {\tt hep-ex/0403026}, submitted to Phys. Rev. Lett.
\bibitem{ref:babar_etapks}
  B.~Aubert {\it et al.} (BaBar Collaboration),
  Phys. Rev. Lett. {\bf 91} 161801 (2003).
\bibitem{ref:smpollution}
  Y.~Grossman, G.~Isidori and M.~P.~Worah, 
  Phys. Rev. D {\bf 58}, 057504 (1998);
  D.~London and A.~Soni, Phys. Lett B {\bf 407}, 61 (1997).
\bibitem{ref:newphys}
  Y.~Grossman and M.~Worah, Phys. Lett. B {\bf 395}, 241 (1997).
\bibitem{ref:babar_kspi0}
  B.~Aubert {\it et al.} (BaBar Collaboration),
  {\tt hep-ex/0403001}, submitted to Phys. Rev. Lett.
\bibitem{ref:bigicartersanda}
  A.~B.~Carter and A.~I.~Sanda, Phys. Rev. D {\bf 23}, 1567 (1981);
  I.~I.~Bigi and A.~I.~Sanda, Nucl. Phys. B {\bf 193}, 85 (1981).
\bibitem{ref:atwood_gronau_soni}
  D.~Atwood, M.~Gronau and A.~Soni,
  Phys. Rev. Lett. {\bf 79}, 185 (1997).
\bibitem{ref:kpipigamma}
  S.~Nishida, M.~Nakao {\it et al.} (Belle Collaboration),
  Phys. Rev. Lett. {\bf 89}, 231801 (2002);
  B.~Aubert {\it et al.} (BaBar Collaboration),
  {\tt hep-ex/0308021}, BABAR-CONF-03/025.
\bibitem{ref:dqstl}
  These modes are noted as a special case of the 
  three body transversity analysis in
  I.~Dunietz {\it et al.},
  Phys. Rev. D {\bf 43}, 2193 (1991).
\bibitem{ref:ddk}
  T.~E.~Browder {\it et al.}, Phys. Rev. D {\bf 61}, 054009 (2000).
\bibitem{ref:dcppi0}
  R.~Fleischer, Phys. Lett. B {\bf 562}, 234 (2003).
\bibitem{belle_dkk}
  A.~Drutskoy {\it et al.} (Belle Collaboration),
  Phys. Lett. B {\bf 542}, 171 (2002).
\bibitem{ref:kkk}
  A.~Garmash, {\it et al.} (Belle Collaboration), 
  Phys. Rev. D. {\bf 69}, 012001 (2004).
\bibitem{ref:resolution_nim}
  H.~Tajima {\it et al.}, 
  {\tt hep-ex/0301026}, submitted to Nucl. Instr. and Meth. A.
\bibitem{ref:babar_dd}
  B.~Aubert {\it et al.} (BaBar Collaboration),
  Phys. Rev. Lett. {\bf 90}, 221801 (2003);
  B.~Aubert {\it et al.} (BaBar Collaboration),
  Phys. Rev. Lett. {\bf 91}, 131801 (2003).
\bibitem{ref:superb_loi}
  A.~G.~Akeroyd {\it et al.}, {\tt hep-ex/0406071}.
\bibitem{ref:sin2phi1_140fb}
  K.~Abe, {\it et al.} (Belle Collaboration), 
  {\tt hep-ex/0308036}, BELLE-CONF-0353.
\bibitem{ref:chic0k}
  K.~Abe {\it et al.} (Belle Collaboration),
  Phys. Rev. Lett. {\bf 88}, 031802 (2002).
\bibitem{ref:pdg} 
  K.~Hagiwara {\it et al.}, 
  Phys. Rev. D {\bf 66}, 010001 (2002).
\bibitem{ref:hazumi_phiphik}
  M.~Hazumi, Phys. Lett. B {\bf 583}, 285 (2004).
\bibitem{ref:phiphik}
  H.-C.~Huang, {\it et al.} (Belle Collaboration),
  Phys. Rev. Lett. {\bf 91}, 241802 (2003).
\bibitem{ref:belle-conf-0338}
  K.~Abe {\it et al.} (Belle Collaboration), BELLE-CONF-0338.
  Similar conclusions may be drawn from
  B.~Aubert {\it et al.}  (BaBar Collaboration),
  {\tt hep-ex/0308065}, submitted to Phys. Rev. D.
\bibitem{ref:pi0pi0}
  S.~H.~Lee, K.~Suzuki {\it et al.} (Belle Collaboration),
  Phys. Rev. Lett. {\bf 91}, 261801 (2003);
  B.~Aubert {\it et al.} (BaBar Collaboration),
  Phys. Rev. Lett. {\bf 91}, 241801 (2003).
\bibitem{ref:snyder-quinn}
  A.~E.~Snyder and H.~R.~Quinn, Phys. Rev. D {\bf 48}, 2139 (1993).
\bibitem{ref:pipipi0}
  A.~Deandrea and A.~D.~Polosa, Phys. Rev. Lett {\bf 86}, 216 (2001);
  S.~Gardner and U.-G.~Mei\ss ner, Phys. Rev. D {\bf 65} 094004 (2002).
\bibitem{ref:rhopi}
  A.~Gordon, Y.~Chao {\it et al.} (Belle Collaboration),
  Phys. Lett. B {\bf 542}, 183 (2002);
  B.~Aubert {\it et al.} (BaBar Collaboration),
  Phys. Rev. Lett. {\bf 91}, 201802 (2003);
  B.~Aubert {\it et al.} (BaBar Collaboration),
  {\tt hep-ex/0311049}, submitted to Phys. Rev. Lett.
\bibitem{ref:rho0pi0}
  J.~Dragic, T.~Gershon {\it et al.} (Belle Collaboaration),
  {\tt hep-ex/0405068}, submitted to Phys. Rev. Lett.

\end{thebibliography}
\end{document}